\documentclass[11pt]{cernrep}
\usepackage{graphics,epsfig}
\usepackage{cite}

\def\dalemb#1#2{{\vbox{\hrule height .#2pt
        \hbox{\vrule width.#2pt height#1pt \kern#1pt
                \vrule width.#2pt}
        \hrule height.#2pt}}}

 \def\bd{\begin{document}} \def\ed{\end{document}}
\def\ds{\documentstyle} \let\fr=\frac \let\bl=\bigl \let\br=\bigr
\def\citere#1{\mbox{Ref.~\cite{#1}}}
\def\citeres#1{\mbox{Refs.~\cite{#1}}}
\let\Br=\Bigr \let\Bl=\Bigl 
\let\bm=\bibitem
\let\na=\nabla
\let\pa=\partial \let\ov=\overline
\def\ie{{\it i.e.\ }} 
\def\beq{\begin{equation}}
\def\eeq{\end{equation}}
\def\beqa{\begin{eqnarray}}
\def\eeqa{\end{eqnarray}}

\newcommand{\ba}{\begin{array}}
\newcommand{\ea}{\end{array}}
\newcommand{\td}{\tilde}
\newcommand{\norsl}{\normalsize\sl}
\newcommand{\ns}{\normalsize}
\newcommand{\refs}[1]{(\ref{#1})}
\def\simlt{\mathrel{\lower2.5pt\vbox{\lineskip=0pt\baselineskip=0pt
           \hbox{$<$}\hbox{$\sim$}}}}
\def\simgt{\mathrel{\lower2.5pt\vbox{\lineskip=0pt\baselineskip=0pt
           \hbox{$>$}\hbox{$\sim$}}}}

\def\mua{\marginpar[\boldmath\hfil$\uparrow$]%
                   {\boldmath$\uparrow$\hfil}%
                    \typeout{marginpar: $\uparrow$}\ignorespaces}
\def\mda{\marginpar[\boldmath\hfil$\downarrow$]%
                   {\boldmath$\downarrow$\hfil}%
                    \typeout{marginpar: $\downarrow$}\ignorespaces}
\def\mla{\marginpar[\boldmath\hfil$\rightarrow$]%
                   {\boldmath$\leftarrow $\hfil}%
                    \typeout{marginpar: $\leftrightarrow$}\ignorespaces}

\begin{document}

\title{Kaluza-Klein states at the LHC in models with localized fermions}
\author{E.~Accomando and K.~Benakli}
\maketitle
\begin{abstract}
We give a brief review of some aspects of physics with TeV size
extra-dimensions. We focus on a minimal model with matter localized at the 
boundaries
for the study of the production of Kaluza-Klein excitations of gauge bosons.
We briefly discuss different ways to depart from this simple analysis.
\end{abstract} 

\section{INTRODUCTION}
Despite the remarkable success of the Standard Model (SM) in describing the 
physical phenomena at the energies probed at present accelerators, some of 
its theoretical aspects are still unsatisfactory. One of the lacking parts
concerns understanding the gravitational interactions as they  destroy the 
renormalizability of the theory. Furthermore, these quantum gravity effects 
seem to imply the existence of extended objects living in more than four 
dimensions. This raises many questions, as:

Is it possible that our world has more dimensions than those we are aware of?
If so, why don't we see the other dimensions? Is there a way to detect them? 

Of course, the answer to the last question can only come for specific class
of models as it depends on the details of the realization of the 
extra-dimensions and the way known particles emerge inside them.
The examples discussed in this review are the pioneer models described in 
Refs. \cite{Ant, pheno1,pheno2, pheno3, pheno4}, when embedded in the complete
and consistent framework given in \cite{ADD,AADD1}.
We focus on such a scenario as our aim is to understand the most 
important concepts underlying extra-dimension physics, and not to 
display a collection of hypothetical models.  
  
Within our framework, two fundamental energy scales play a major role. The 
first one, $M_s=l_s^{-1}$, is related to the inner structure of the basic 
objects of the theory, that we assume to be elementary strings. 
Their point-like behavior is viewed as a low-energy phenomena; above $M_s$, 
the string oscillation modes get excited making their true extended nature 
manifest.
The second important scale, $R^{-1}$, is associated with the existence of a 
higher dimensional space. Above $R^{-1}$ new dimensions open up and 
particles, called Kaluza-Klein (KK) excitations, can propagate in them.

\begin{figure}
\begin{center}
\includegraphics[width=0.5\textwidth]{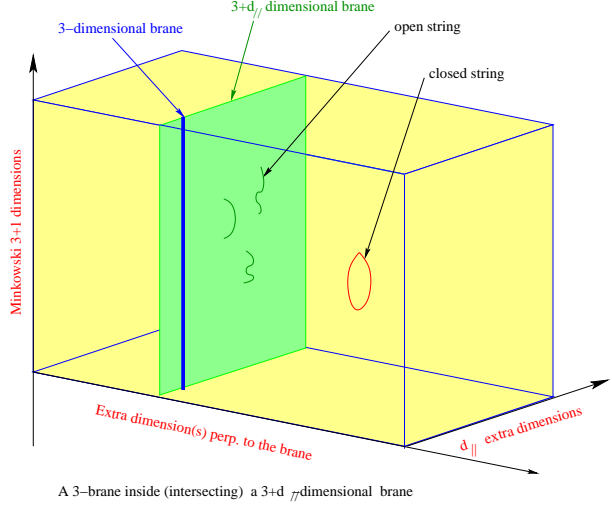}
\caption{Geometrical representation of models with localized fermions.}
\label{fig:fig1}
\end{center}
\end{figure}

\section{MINIMAL MODELS WITH LOCALIZED FERMIONS}

In a pictorial way, gravitons and SM particles can be represented as
in Fig. 1. In particular, in the scenario we consider:
\begin{itemize}
\item the gravitons, depicted as closed strings, are seen to propagate in the 
whole higher-dimensional space, 3+$d_\parallel$+$d_\perp$. Here, 
3+$d_\parallel$ defines the longitudinal dimension of the big brane drawn in 
Fig. 1, which 
contains the small 3-dimensional brane where the observed SM particles live.
The symbol $d_\perp$ indicates instead the extra-dimensions, transverse to
the big brane, which are felt only by gravity.
\item The SM gauge-bosons, drawn as open strings, can propagate only on 
the (3+$d_\parallel$)-brane. 
\item The SM fermions are localized on the 3-dimensional brane, which 
intersects the (3+$d_\parallel$)-dimensional one. They do not propagate on
extra-dimensions (neither $d_\parallel$ nor $d_\perp$), hence they do not 
have KK-excitations.
\end{itemize}

\noindent
The number of extra-dimensions, 
$D=d_\parallel$,$d_\perp$ or $d_\parallel$+$d_\perp$, which are compactified 
on a $D$-dimensional torus of volume $V=(2 \pi)^D  R_1 R_2 \cdots R_D$, can 
be as big as six \cite{AADD1} or seven \cite{karim} dimensions. Assuming 
periodic conditions on the wave functions along each compact direction, the 
states propagating in the $(4+D)$-dimensional space are seen from the 
four-dimensional point of view as a tower of states having a squared mass:
\begin{equation}
M^2_{KK}\equiv M^2_{\vec n} = m_0^2 +\frac {n_1^2}{R_1^2} +
\frac {n_2^2}{R_2^2}+ \cdots  +\frac {n_D^2}{R_D^2}\, ,
\label{KKdef}
\end{equation}
with $m_0$ the four-dimensional mass and $n_i$ non-negative integers. The 
states with $\sum_i n_i \neq 0$ are called KK-states. Assuming that leptons 
and quarks are localized is quite a distinctive feature of this class of 
models, giving rise to well defined predictions. An immediate  consequence of 
the localization is that fermion interactions do not preserve the momenta in 
the extra-dimensions. One can thus produce single KK-excitations, for example 
via $f\bar{f^\prime}\rightarrow V_{KK}^{(n)}$ where $f,f^\prime$ are fermions 
and $V_{KK}^{(n)}$ represents massive KK-excitations of $W,Z,\gamma ,g$ 
gauge-bosons. Conversely, gauge-boson interactions conserve the internal 
momenta, making the self-interactions of the kind 
$VV\rightarrow V_{KK}^{(n)}$ forbidden. The experimental bounds on 
KK-particles that we summarize in the following, as well as the discovery 
potential of the LHC, depend very sensitively on the assumptions made.

Electroweak measurements can place significant limits on the size of the 
extra-dimensions. KK-excitations might affect low-energy observables through 
loops. Their mass can thus be constrained by fits to the electroweak 
precision data \cite{ewdata1,ewdata2,ewdata3,ewdata4,ewdata5,ewdata6,ewdata7}.
In particular, the fit to the measured values of $M_W$, $\Gamma_{ll}$ and 
$\Gamma_{had}$ has led to $R^{-1}\ge$ 3.6 TeV. 

\section{WHAT CAN BE EXPECTED FROM THE LHC?}
 
The possibility to produce gauge-boson KK-excitations at future colliders was 
first suggested in Ref. \cite{pheno3}. Unfortunately, from the above-mentioned
limits, the discovery at the upgraded Tevatron is already excluded 
(see for instance \cite{AAB}). Also expectations of a spectacular explosion 
of new resonances at the LHC are sorely disappointed. In the most optimistic  
case, the LHC will discover just the first excitation modes. 

The only distinctive key from other possible non-standard models with new 
gauge-bosons would be the almost identical mass of the KK-resonances of all
gauge bosons. Additional informations would
be however needed to bring clear evidence for the higher-dimensional origin of 
the observed particle. Despite the interpretation difficulties, detecting a 
resonance would be of great impact. 

We could also be in the less favorable case in which the mass of the 
KK-particles is bigger than the energy-scale probed at the LHC. In this 
unfortunate but likely scenario, the indirect effect of such particles would 
only consists in a slight increase of the events at high energies compared to 
the SM predictions. In this case, the luminosity plays a crucial role. In the 
last few years, several analysis have been performed in order to estimate the 
possible reach of the LHC (see for example \cite{pheno3,AAB,lhcpheno6,lhcpheno5,
pheno12,lhcpheno1,deruj,lhcpheno2,lhcpheno3}).

The three classes of processes where the new KK-resonances could be observed 
are:
\begin{itemize}
\item $pp\rightarrow l^+l^-$,
\item $pp\rightarrow l\nu_l$, where $l\nu_l$ is for $l^+\nu_l+\bar\nu_ll^-$,
\item $pp\rightarrow q\bar q$, where $q=u,d,s,c,b$.
\end{itemize}
The first class can be mediated by the KK-excitations of the electroweak 
neutral gauge-bosons, $Z_{KK}^{(n)}$ and $\gamma_{KK}^{(n)}$, while the 
second one can contain the charged $W_{KK}^{(n)}$ gauge-boson modes. Finally, 
the third class can receive contributions from all electroweak gauge-bosons 
plus the KK-modes $g_{KK}^{(n)}$ of the gluons. 
\vskip -4.cm
\begin{figure}
\begin{center}
\includegraphics[width=0.5\textwidth]{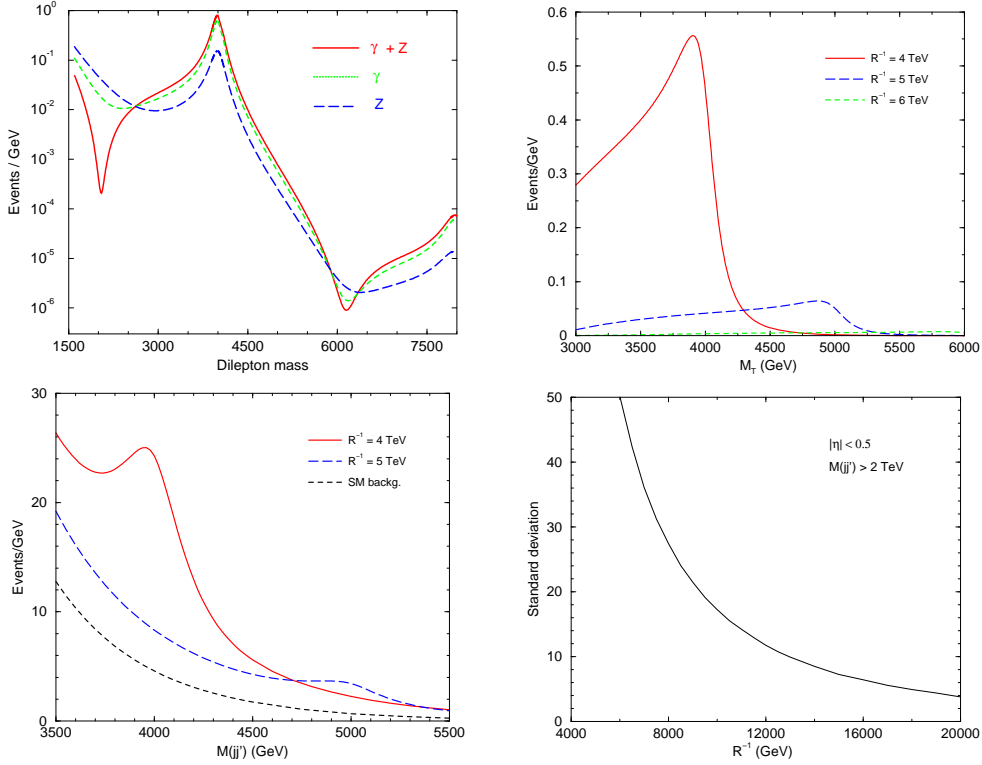}
\vskip -2.5cm
\caption{(a) Resonances of the first KK-excitation modes of $Z$ and $\gamma$
gauge-bosons. (b) Resonances of the first KK-excitation mode of the $W$-boson.
(c) Resonances of the first KK-excitation mode of the gluon. (d) 
Under-hreshold effects due to the presence of $g_{KK}^{(n)}$, given in terms
of the number of standard deviations from the SM predictions.
The results have been obtained for the LHC with $\sqrt{s}$=14 TeV and L=100 
$fb^{-1}$.}
\label{fi:fig2}
\end{center}
\end{figure}
\vskip 4.cm
Typically, one can expect a kind of signal as given in Fig. 2. 
In the case where both outgoing particles are visible, a natural observable is
the invariant mass of the fermion pair. Distributions in such a variable are 
shown in the upper and lower left-side plots, which display the
interplay between $Z_{KK}^{(n)}$ and $\gamma_{KK}^{(n)}$ resonances, and 
the peaking structure due to $g_{KK}^{(n)}$, respectively.
In presence of a neutrino in the final state, one can resort to the transverse
mass distribution in order to detect new resonances. This is shown in the
upper right-side plot of Fig. 2 for the charged-current process with 
$W_{KK}^{(n)}$ exchange.
Owing to the PDFs, the effective center-of-mass (CM) energy of the partonic 
processes available at the LHC is not really high. The discovery limits of 
the KK-resonances are thus rather modest, $R^{-1}\le$ 5-6 TeV. This estimate 
finds confirmation in more detailed ATLAS and CMS analyses \cite{atlascms}.
Taking into account the present experimental bounds, there is no much space 
left. Moreover, the resonances due to the gluon excitations have quite large 
widths owing to the strong coupling value. They are thus spread and difficult 
to detect already for compactification scales of the order of 5 TeV.

But, what represents a weakness in this context can become important for 
indirect searches. The large width, ranging between the order of a few 
hundreds GeV for the KK-excitations of the electroweak gauge-bosons and the
TeV-order for the KK-modes of the gluons, can give rise to sizeable 
effects even if the mass of the new particles is larger than the typical
CM-energy available at the LHC. This is illustrated in the lower right-side 
plot of Fig. 2, where the number of standard deviations quantifies the
discrepancy with the SM predictions, coming from $g_{KK}^{(n)}$ contributions.
The under-threshold effects are driven by the tail of the broad Breit-Wigner, 
which can extend over a region of several TeV, and 
are dominated by the interference between SM and KK amplitudes. They thus 
require to have non-suppressed SM contributions.    
Their size, of a few-per-cent order for large compactification radii, can 
become statistically significant according to the available luminosity. 
In the extreme case of Fig. 2, we have a KK-gluon with mass 
$M_1=R^{-1}$=20 TeV and width $\Gamma_1\simeq$2 TeV. Assuming a luminosity
L=100$fb^{-1}$, the interference terms give rise to an excess of about 2000 
events. Similar conclusions hold for the indirect search of the 
KK-excitations of the electroweak gauge-bosons. At 95\% confidence level, the 
LHC could exclude values of compactification scales up to 12 and 14 TeV from 
the $Z_{KK}^{(n)}+\gamma_{KK}^{(n)}$ and $W_{KK}^{(n)}$ channels, 
respectively. The indirect search is exploited in the ATLAS and CMS joint
analysis of Ref. \cite{atlascms}. 

\section{GOING BEYOND MINIMAL}

We have carried the discussion above for the case of one extra-dimension with 
all fermions localized on the boundaries. 
One can depart from this simple situation in many ways:
\begin{itemize}
\item {\it {\bf More extra-dimensions}}

New difficulties arise for $D \geq 2$: the sum over KK propagators diverges 
\cite{pheno1}.
A simple  regularization is to cut off the sum of the KK states at $M_s$.
This would be natural if the extra-dimension were discrete, however in our 
model we assumed translation invariance of the background geometry (before 
localizing any objects in it). String theory seems to choose a different 
regularization \cite{pheno1,Antoniadis:2000jv}. In fact the interaction of 
$A^\mu (x,\vec y)=\sum_{{\vec n}}
{\cal A}^{\mu }_{\vec n}(x) \exp{i\frac {n_i y_i}{R_i}}$ with the current
density $j_\mu (x)$, associated to the massless localized fermions, is 
described by the effective Lagrangian: 
\begin{eqnarray} 
\int d^4x \, \, \,  \, \sum_{{\vec
n}} e^{-\ln {\delta} \sum_i\frac{n_i^2l_s^2}{2 R_i^2}} \, \,
\, \, \, j_\mu (x) \, {\cal A}^{\mu }_{\vec n}(x)\, , 
\end{eqnarray} 
which can be written after Fourier transformation as 
\begin{eqnarray} \int d^{4}y \,\int d^4x  \, \,
\, \,  (\frac{1}{l_s^2 2 \pi \ln {\delta}})^{2} e^{- \frac {{\vec
y}^2}{2 l_s^2 \ln {\delta}}}  \, j_\mu (x) \, A^\mu (x,\vec y)\, .
\label{brwidth}
\end{eqnarray}
This means that the localized fermions are felt as a Gaussian distribution of 
charge 
$e^{-\frac {{\vec y}^2}{2 \sigma^2}}
 j_\mu (x)$  with a width $\sigma=\sqrt{\ln {\delta}}\, l_s \sim 1.66 \, l_s$. 
Here we used $\delta=16$ corresponding to a $Z_2$ orbifolding. The couplings 
of the massive KK-excitations to the localized fermions are then given by:
\begin{eqnarray}
g_{{\vec n}} = {\sqrt{2}}\sum_{{\vec
n}} e^{-\ln {\delta} \sum_i\frac{n_i^2l_s^2}{2 R_i^2}} g_0
\label{coupling}
\end{eqnarray}
where the factor ${\sqrt{2}}$ stands for the relative normalization of the 
massive KK wave function  ($\cos(\frac {n_i y_i}{R_i})$) with respect to the 
zero mode, and $g_0$ represents the coupling of the corresponding SM 
gauge-boson.

The amplitudes depend on both $R$ and $M_s$ and thus, as phenomenological 
consequence, all bounds depend on both parameters (see \cite{AAB}).

\item {\it {\bf Localized kinetic and/or mass terms for bulk fields}}

Let us denote by $S_0(p,R,M_s)$ the sum of all tree-level boson propagators 
weighted by a factor $\delta^{-\frac {{\vec n}^2}{R^2 M_s^2}}$ from the interaction 
vertices. For simplicity we take $m_0=0$, and define $\delta S_0$ by
\begin{equation}
S_0(p,R,M_s)= {1 \over p^2} + \delta S_0 \  .
\label{Sbar}
\end{equation}

In order to confront the theory with experiment, it is necessary to include 
a certain number of corrections. The obvious one is a resummation of one-loop
self-energy correction to reproduce the gauge coupling of the massless 
vector-bosons.
Here we parametrize these effects as two kinds of bubbles to be resummed:

\begin{itemize}
\item the first, denoted as ${\cal B}_{bulk}$ represents the bulk 
corrections. This bubble preserves the KK-momentum,
\item the second, denoted as ${\cal B}_{bdary}$ represents the boundary 
corrections. This bubble does not preserve the KK momentum. In fact, this can 
represent a boundary mass term or tree-level coupling, but also localized
one-loop corrections due to boundary states \cite{pheno1} or 
induced by bulk states themselves \cite{Georgi:2000ks}.
\end{itemize}

Here, two simplifications have been made: (a) the corrections are the same 
for all KK-states, and (b) the boundary corrections arise all from the same 
boundary. This results in the corrected propagator  \cite{pheno1}: 
\begin{equation}
S_{corr}(p,R,M_s)= {S_0 \over {1- {\cal B}_{bulk} -{\cal B}_{bdary} -
 p^2 \delta S_0 {\cal B}_{bdary}}}\ .
\label{Sren}
\end{equation}
If we define the ``renormalized coupling'' as
 $g^2(p^2) ={ g^2 \over {1- {\cal B}_{bulk} -{\cal B}_{bdary} }}$, the result is
\begin{equation}
g^2 S_{corr}= g^2(p^2)S_0 - \delta S_0 \frac{g^2 (1- p^2 \delta S_0) {\cal B}_{bdary}}
{(1- {\cal B}_{bulk} -{\cal B}_{bdary})(1- {\cal B}_{bulk} -{\cal B}_{bdary}p^2 \delta S_0)} \ .
\label{Srenf}
\end{equation}
The first term in Eq.(\ref{Srenf}) is the contribution that was taken into 
account in all phenomenological analysis, the second is the correction which 
depends crucially on the size of ${\cal B}_{bdary}$.

\item {\it {\bf Spreading interactions in the extra dimensions}}

In the simplest scenario, all SM gauge-bosons propagate in the same compact 
space. However, one may think that the three factors of the SM gauge-group
can arise from different branes, extended in different compact directions.
In this case, $d_\parallel$ TeV-dimensions might be longitudinal to some 
brane and transverse to others. As a result, only some of the gauge-bosons 
can exhibit KK-excitations. Such a framework is discussed in 
\cite{AAB}.
\end{itemize}
\noindent
These are simplest extensions of the work we presented above. The 
experimental limits depend now on many parameters
$M_s, {\cal B}_{bdary}, ...$ in addition to the different size of the 
compactification space felt by the gauge-bosons. 
   
\bibliography{accom4}
\end{document}